\documentclass[%
 reprint,
 amsmath,amssymb,
 aps,
]{revtex4-1}

\usepackage{graphicx}
\usepackage{dcolumn}
\usepackage{bm}
\usepackage{hyperref}

\usepackage{csquotes} 

\usepackage{xcolor}

\providecommand\diffd{{\mathrm d}}
\providecommand\diffD{{\mathrm D}}
\providecommand\Sect{\S}

\providecommand\kmsMpc{km~s$^{-1}$~Mpc$^{-1}$}

\begin{document}

\title{Cosmological signatures of torsion and how to distinguish torsion from the dark sector}

\author{Krzysztof Bolejko}
 \email{krzysztof.bolejko@utas.edu.au}
\affiliation{School of Natural Sciences, College of Sciences and Engineering, University of Tasmania, Private Bag 37, Hobart TAS 7001, Australia}

\author{Matteo Cinus, Boudewijn F. Roukema}
\affiliation{Institute of Astronomy, Faculty of Physics, Astronomy and Informatics,
  Nicolaus Copernicus University, Grudziadzka 5,
  87-100 Toru\'n, Poland}


\begin{abstract}
Torsion is a non-Riemannian geometrical extension of general relativity that allows including the spin of matter and the twisting of spacetime. Cosmological models with torsion have been considered in the literature to solve problems of either the very early (high redshift $z$) or the present-day Universe. This paper focuses on distinguishable observational signatures of torsion that could not be otherwise explained with a scalar field in pseudo-Riemannian geometry.
We show that when torsion is present, the cosmic duality relation between the angular diameter distance, $D_{\mathrm A}$, and
the luminosity distance, $D_{\mathrm L}$, is broken. We show how the deviation described by the parameter $\eta = D_{\mathrm L}/[ D_{\mathrm A}(1+z)^2] -1 $ is linked to torsion and how different forms  of torsion lead to special-case parametrisations of $\eta$, including $\eta_0 z$, $\eta_0 z/(1+z)$, and $\eta_0 \ln (1+z)$.
We also show that the effects of torsion could be visible in low-redshift data, inducing biases in supernovae-based $H_0$ measurements. We also show that torsion can impact the Clarkson--Bassett--Lu (CBL) function ${\cal C}(z) = 1 + H^2 (D D'' - D'^2) + H H' D D'$, where $D$ is the transverse comoving distance. If $D$ is inferred from the luminosity distance, then, in general non-zero torsion models, ${\cal C}(z) \ne 0$. For pseudo-Riemannian geometry, the
Friedmann--Lema{\^\i}tre--Robertson--Walker (FLRW) metric has ${\cal C}(z) \equiv 0$; thus, measurement of the CBL function could provide another diagnostic of torsion.
\end{abstract}

\maketitle

\section{Introduction}

General relativity (GR) describes gravitational interaction in terms of spacetime curvature. The theory is built upon pseudo-Riemannian geometry (hereafter, \enquote*{Riemannian} geometry implicitly refers to Lorentzian geometry). Within Riemannian geometry, the motion of free particles (i.e. no external forces) is fully determined by the Riemann curvature tensor.
Moving beyond GR one finds that curvature is not the only geometrical feature that could affect physical processes.
One such geometrical feature is torsion. Torsion could be generated by a spin tensor for
matter, in addition to the energy--momentum tensor of GR (e.g., \cite[][]{2006gr.qc.....6062T}).  An
example of a theory with torsion is the Einstein--Cartan theory of gravity \cite{2006gr.qc.....6062T}.
A very special case of models with torsion is that of models with vanishing curvature ($R=0$ and so $g_{\alpha \beta} \equiv \eta_{\alpha \beta}$, where $\eta_{\alpha \beta}$ is the Minkowski metric). When curvature vanishes, the connection reduces to the
Weitzenb\"ock connection, which is then specified by torsion alone. Theories of this type
include those whose field equations can be obtained from the Einstein--Hilbert action
proportional to torsion $T$, called $f(T)$ models,
including \enquote{teleparallel gravity} for the case where $f(T)=T$
\cite[][\Sect4.2.5]{2012PhR...513....1C}.

Torsion allows for additional degrees of freedom, and so in the past, models with torsion were proposed to solve certain early universe problems, replacing the initial singularity with a big bounce and inflation \cite{2012PhRvD..85j7502P,2017EL....11929002A,Cubero_2019}. Recently, cosmological models with torsion were used to investigate alternatives to dark energy \cite{2019EPJC...79..950P,2020PDU....2700416M}.
Models with torsion exhibit many properties of models with scalar fields. Thus, from the cosmological point of view, it is reasonable to ask questions such as: is torsion just a geometrical replacement for a scalar field, and if it were, would it survive the scrutiny of Ockham's razor?
On the other hand, if there are cosmological signatures of torsion that are distinguishable from those of dark sector physics, then what are they, and could we use cosmological data to distinguish between dark energy and torsion?

This paper focuses on light propagation and aims at discussing the observational signatures of non-Riemannian effects on cosmological scales. The aim of this paper is to identify
those cosmological signatures that would require models with torsion,
i.e., signatures that could be not be produced by dark energy models
but would require a departure from Riemannian geometry.
In addition to non-zero torsion, geometrical spacetime models can also be extended
beyond general relativity to allow non-metricity \cite[e.g.][]{MaoTGC07}. We refer briefly
to these for completeness.

\section{Metric-affine space}
A metric-affine space is characterised by a
metric ${\mathbf g}$ and a linear connection ${\Gamma}$ \cite{1995PhR...258....1H}.
The metric and the connection are independent objects. The metric ${\mathbf g}$
can be used for evaluating the magnitude of vectors
\begin{equation}
v=\sqrt{g_{\alpha\beta}\,v^\alpha v^\beta} ,
\end{equation}
and the connection ${\Gamma}$
determines the transport of
vectors $v^\alpha$ along a curve $C$ with a tangent vector $t^\sigma$,
with parallel transport of the projection of $v^\alpha$ on $t^\sigma$:
\begin{equation}
t^\sigma \nabla_\sigma v^\alpha = t^\sigma \partial_\sigma v^\alpha
+ t^\sigma\Gamma^\alpha_{\nu \sigma}  v^\nu = 0.
\end{equation}
The connection coefficient can be written as \cite{Fabbri2006}
\begin{equation}
 \Gamma^\sigma_{\alpha \beta} =\left\{{\sigma \atop \alpha\beta}\right\} +L^\sigma_{\phantom{\sigma}\alpha \beta} +
K^\sigma_{\phantom{\sigma}\alpha \beta},
\label{connection}
\end{equation}
 where $\left\{{\sigma \atop \alpha\beta}\right\}$ are the Christoffel symbols:
\begin{equation}
 \left\{{\sigma \atop \alpha\beta}\right\}
 = \frac{1}{2} g^{\sigma \nu} \left( \partial_\beta g_{\nu \alpha} +  \partial_\alpha g_{\nu \beta}
 - \partial_\nu g_{\alpha \beta} \right),
 \label{e-Christ-sym-vs-metric}
\end{equation}
and $L^\sigma_{\phantom{\sigma}\alpha\beta}$ is the metric incompatibility tensor
\begin{equation}
 L^\sigma_{\phantom{\sigma}\alpha \beta}
 = \frac{1}{2} g^{\sigma \nu} \left( \nabla_\beta g_{\nu \alpha} +  \nabla_\alpha g_{\nu \beta}
 - \nabla_\nu g_{\alpha \beta} \right),
  \label{e-L-vs-metric}
\end{equation}
where ${\mathbf \nabla}$ is a covariant derivative with respect to the connection
$\Gamma^\sigma_{\alpha \beta}$ defined in \eqref{connection}.
While the Levi-Civita connection is fully determined by the metric itself (and its partial derivatives) to evaluate components of $L^\sigma_{\phantom{\sigma}\alpha \beta}$ one needs to specify the connection first. The contorsion tensor $K^\sigma{}_{\alpha \beta}$ is defined in terms of the torsion tensor, which describes the fact that the connection is not symmetric in its lower indices
 \begin{equation}
T^\sigma_{\phantom{\sigma}\alpha \beta}{} := \Gamma^\sigma_{\alpha \beta} - \Gamma^\sigma_{\beta \alpha }.
\label{torsion}
\end{equation}

If one requires that the connection is \enquote*{metric} then
   \begin{equation}
\nabla_\sigma g_{\alpha \beta}  = 0.
 \end{equation}
 The metricity condition imposes that $L^\sigma_{\phantom{\sigma}\alpha \beta} = 0$,
 and the difference between the affine connection
and the Levi-Civita connection is described by the contorsion tensor alone.
In the literature, there are various conventions regarding the contorsion tensor;
here we follow \cite[][(12)]{2006gr.qc.....6062T})
 \begin{equation}
 K_{\sigma \alpha \beta}=
 \frac{1}{2} \left( T_{  \sigma \alpha  \beta }  +  T_{   \beta \sigma \alpha }+T_{  \alpha \sigma  \beta }   \right).
 \label{contorsion}
\end{equation}
 The contorsion tensor is antisymmetric in the first and third indices
   $  K_{\sigma \alpha \beta} + K_{\beta \alpha \sigma} = 0$.

\subsection{Weak Equivalence Principle and torsion}

In general, no constraint is required on the last two indices of contorsion, i.e. relating $K_{\sigma\alpha\beta}$
to $K_{\sigma\beta\alpha}$,
which means that
the symmetric part of the connection does, in general, contain some combination of the torsion tensor, and thus
\begin{equation*}
\Gamma^\sigma_{(\alpha \beta)}  \ne \left\{ {\sigma \atop \alpha\beta} \right\}.\end{equation*}
This means that the extremal curves, i.e. those that satisfy
   \begin{equation}
 \frac{\mathrm{d}^2x^\alpha}{\mathrm{d}s^2}+
 \left\{ {\alpha \atop \beta\gamma} \right\}
 \frac{\mathrm{d}x^\beta}{\mathrm{d}s}\frac{\mathrm{d}x^\gamma}{\mathrm{d}s}=0,
   \end{equation}
   and the autoparallel curves
      \begin{equation}
 \frac{\mathrm{d}^2x^\alpha}{\mathrm{d}s^2}+\Gamma^\alpha_{(\beta \gamma)}\frac{\mathrm{d}x^\beta}{\mathrm{d}s}\frac{\mathrm{d}x^\gamma}{\mathrm{d}s}=0,
   \end{equation}
need not be the same.
If one requires that extremal curves are also autoparallel curves then this puts a constraint
on torsion. This condition is required for a freely-falling frame to exist in the
vicinity of a point O, with
\begin{eqnarray}
 \left. g_{\alpha\beta} \right|_O = \eta_{\alpha\beta}, \nonumber \\
 \left. \partial_\nu g_{\alpha\beta} \right|_O = 0, \nonumber \\
 \left. \Gamma^\sigma_{(\alpha \beta)} \right|_O = 0. \nonumber
\end{eqnarray}
If one requires the Weak Equivalence Principle to hold in the presence of torsion, then the
 torsion tensor is totally antisymmetric in all of its indices \cite{Yu1989}
\begin{equation}
    T_{\alpha \beta\sigma} = T_{[\alpha \beta]\sigma} = T_{\alpha [\beta\sigma]}.
\end{equation}
Here, we do not assume this condition; instead, we allow extremal curves to differ from autoparallel ones. As shown below (and represented in Fig. \ref{fig-nullcurve}), when torsion is present, the null curves do not need to be extremal or autoparallel \cite{2017PhRvD..95f1501S}.

\section{Light propagation}

\subsection{Geometric optics approximation}\label{geoopt}

Assuming the geometric optics approximation, the electromagnetic wave can be approximated as \cite{1966ApJ...143..379K,LL1987,Ellis2009,PK2006}

\begin{equation}
F = a(x^\alpha) g(\varphi),
\label{EMansatz}
\end{equation}
where $a$ is the amplitude of the wave that depends on space-time coordinate $x^\alpha$,
$g$ is an arbitrary smooth function, and $\varphi$ the
eikonal \cite[the phase,][]{LL1987}. The wave vector $k^\alpha$ is defined as a gradient of the eikonal
\begin{equation}
   k^\alpha = g^{\alpha \beta} \,\nabla_\beta \,\varphi.
   \label{nullvector}
\end{equation}
The wave vector is a vector tangent to the light curve, and thus by the nature of the null curve, it must be null as well:
\begin{equation}
  k_\alpha k^\alpha  = 0.
  \label{e-null-curve}
\end{equation}
This is the condition that represents the underlying assumption of light propagation, i.e. that light propagates along null curves.
Inserting the ans{\"a}tz (\ref{EMansatz}) into Maxwell's
equation
yields \eqref{e-null-curve}, confirming that the ans{\"a}tz behaves as
expected \footnote{\protect\mbox{If} the Maxwell equations do not require the light to propagate along the null curves, then either the Maxwell equations or the ansatz would need to be modified.}.
Generalising Maxwell's equation beyond GR is not a trivial and unambiguous task \cite{2015BrJPh..45..353F}. Following the approach based on exterior calculus \cite{Vandyck_1996},
\begin{equation}
{\rm d} F = 0 \Leftrightarrow
    \nabla_{[\alpha} F_{\mu \nu]} + F^\beta{}_{[\mu}T_{|\beta|\nu\alpha]}= 0,
\end{equation}
and using the ans{\"a}tz (\ref{EMansatz}) and following the perturbative approach
\cite{1966ApJ...143..379K,PK2006,Ellis2009},
\begin{equation}
g_{\mu \nu}  k^\mu k^\nu  = 0,
\label{eikonal}
\end{equation}
and
\begin{equation}
 T_{\mu \nu}^{{\rm rad}} =  A^2 k_\mu k_\nu,
\label{em-rad}
\end{equation}
where $T_{\mu \nu}^{\mathrm{rad}}$ is the energy--momentum tensor of radiation,
$ T_{\mu\nu}^{\mathrm{rad}} = F_{\mu \sigma} F^{\sigma}{}_\nu- \frac{1}{4} F_{\alpha\beta} F^{\alpha \beta}$, unrelated to the torsion $T^{\sigma}_{\phantom{\sigma}\alpha\beta}$. The amplitude $A^2 = A_a A^a$, and
$ F_{\alpha\beta} = g' (A_\alpha k_\beta - A_\beta k_\alpha)$.

\begin{figure}
\begin{center}
\includegraphics[scale=0.3]{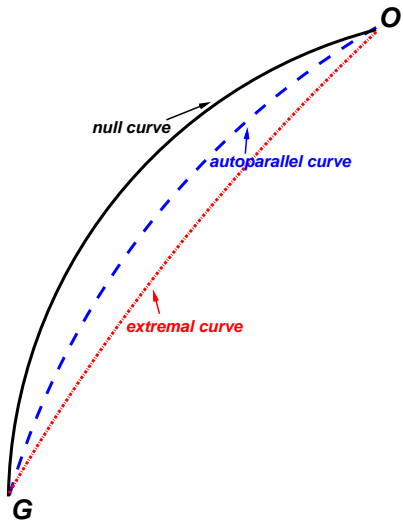}
\end{center}
\caption{A distant galaxy \enquote*{G} emits photons that are observed by observer
  \enquote*{O}, defining two events G and O.
  In GR, joining these two events with
  extremal curves, autoparallel curves or null curves leads to the same result.
  In the presence of torsion, autoparallel curves are no longer extremal;
  moreover, as torsion pushes photons out of geodesics, the null curves are no longer autoparallel.}
\label{fig-nullcurve}
\end{figure}

\subsection{Null curves} \label{s-null-curves}

The eikonal equation (\ref{eikonal}) states that
light propagates along the null curves, hence
\begin{equation}
   \nabla_\beta \left( k_\alpha k^\alpha  \right) = 0 \quad \Rightarrow \quad k_\alpha \, \nabla_\beta  k^\alpha  = 0.
\label{cone}
\end{equation}
If we now apply the commutator of $\nabla$ to the eikonal $\varphi$ and using (\ref{nullvector}), we have
\begin{equation}
 \nabla_\mu k_\nu -  \nabla_\nu k_\mu =  T^\beta_{\phantom{\beta}\mu \nu} \, k_\beta.
\end{equation}
Contracting the above with the null vector $k^\mu$
and using the conservation of the eikonal (\ref{cone}) we obtain the formula for the propagation of the null vector $k^\alpha$
\begin{equation}
g_{\nu \alpha} k^\mu \nabla_\mu k^\alpha =   T_{\alpha  \beta \nu }  \, k^\alpha k^\beta,
\label{propagation}
\end{equation}

In general, the null curve that describes light propagation does not necessarily
have to be autoparallel.
Let us consider three scenarios:

\begin{enumerate}
    \item {\em Totally antisymmetric torsion}:

If torsion is totally antisymmetric $T_{ \alpha  \beta \nu}   \, k^\alpha k^\beta = 0$, hence
\begin{equation}
k^\mu \nabla_\mu k^\alpha = 0,
\end{equation}
which means that light propagates along a geodesic, which is affine parametrised.

\item {\em Torsion aligned with null vector}

If torsion is not totally antisymmetric but the null vector is aligned with the torsion tensor, in the sense that
\begin{equation}
T_{ \alpha  \beta \nu} \, k^\alpha k^\beta = c \, k_\nu
\end{equation}
for a non-zero constant $c$,
then we recover a geodesic again, but this time with a non-affine parametrisation \cite{PK2006}, i.e.
\begin{equation}
g_{\nu \alpha} k^\mu \nabla_\mu k^\alpha =     c \, k_\nu,
\end{equation}
which means that torsion induces inaffinity acceleration but keeps photons on autoparallel curves.
However, in general this can  only hold for some distinct directions
(i.e. those directions that are aligned with the tensor  $T_{ \alpha  \beta \nu}$). An example of such a case is a homogeneous and isotropic FLRW-type model with torsion \cite{Kranas2019}, where the only non-zero elements of the torsion tensors are $T_{i0i}$ (where $i$ is a spatial index); in such a case the distinct direction is time-like and torsion provides a
non-standard redshift contribution at a monopole level.

\item {\em The most general  case}

In the most general case we have
\begin{equation}
\forall c \in \mathbb{R},\; T_{ \alpha \beta \nu } \, k^\alpha k^\beta \ne c \, k_\nu,
\end{equation}
which means that torsion induces an orthogonal acceleration component and modifies the trajectory of photons,
pushing them out of geodesics.
In such a case a light curve is no longer autoparallel \cite{2017PhRvD..95f1501S,2018PhRvD..98h4029S}.
\end{enumerate}

Assuming only that light propagates along null curves, i.e. that the vector
tangent to the light path is null, $k_\alpha k^\alpha = 0$, we obtain
that in the most general case, in a spacetime with torsion,
not only does
light not move along extremal curves, but neither does it move along
geodesics.  Using an (imperfect) analogy with pressure gradients
pushing dust particles out of the (time-like) geodesics, we could say
that in some sense torsion `exerts force' on photons and `pushes' them
out of the geodesics; cf. Fig.~\ref{fig-nullcurve}.

\section{Redshift}

We assume no difference from the usual case in which the time component of a photon's 4-momentum
as measured by an observer with velocity $u^\alpha$ is its energy $E$,
and that this relates to the frequency $\nu$ in the usual quantum mechanical way,
where we write the Planck constant as $h=1$
\begin{equation}
E = \nu = - k^\alpha u_\alpha .
\end{equation}
We can now write the propagation equation to evaluate how frequency changes as light propagates from the emitter to the observer.

\begin{equation}
 \frac{{\rm D} \nu}{{\rm D} s} =  - k^\mu \nabla_\mu ( u_\alpha k^\alpha) =   - k^\alpha k^\mu \nabla_\mu u_\alpha - u_\alpha k^\mu \nabla_\mu k^\alpha.
\end{equation}
Using the decomposition of the velocity gradient onto the scalar of expansion $\Theta$
and shear $\sigma_{\alpha \mu}$ (where $ \nabla_\mu  u_\alpha = \sigma_{\alpha\mu} + h_{\alpha\mu} \Theta/3 $), we obtain
\begin{equation}
 \frac{{\rm D} \nu}{{\rm D} s} =
  - k^\alpha k^\mu \left[ \left( g_{\alpha \mu} + u_\alpha u_\mu  \right) \frac{\Theta}{3}
  + \sigma_{\alpha \mu} +   u^\sigma  T_{ \alpha  \mu \sigma} \right].
  \label{freqprop}
 \end{equation}
Using the decomposition of the null vector into its temporal and spatial parts \cite[][(6.12)]{Ellis2009}
\begin{equation}
    k^\alpha = \nu (u^\alpha + n^\alpha),
    \label{nvdecom}
\end{equation}
the frequency propagation equation (\ref{freqprop}) reduces to
\begin{equation}
 \frac{{\rm D} \nu}{{\rm D} s} = - \nu^2 \left(\frac{1}{3}\Theta + \Sigma + T\right),
 \label{redshiftpropagation}
 \end{equation}
where $\Sigma=\sigma_{\alpha \beta} \, n^\alpha n^\beta$,
and $T = T_{ \alpha  \beta \sigma} \, u^\alpha n^\beta u^\sigma
+ T_{ \alpha  \beta \sigma}\,   n^\alpha n^\beta u^\sigma $.

Changing the time variable via $ {\rm d} t  =  \nu \,{\rm d} s $ \cite{Ellis2009}, we obtain
\begin{equation}
    \ln \left( \frac{\nu_{\mathrm G}}{\nu_{\mathrm O}} \right) = \int\limits_{\mathrm G}^{\mathrm O} {\rm d} t \, \left(\frac{1}{3}\Theta + \Sigma +T \right).
    \label{ltb}
\end{equation}
For homogeneous and isotropic models the right-hand side reduces to $\ln (a_{\mathrm O}/a_{\mathrm G})$,
where $a$ is the scale factor rather than the wave amplitude,
yielding a redshift with no dependence on direction. For inhomogeneous models,
the RHS has, in general, directional dependence, yielding a direction-dependent redshift.

\section{Reciprocity theorem}

\begin{figure}
\begin{center}
\includegraphics[scale=0.3]{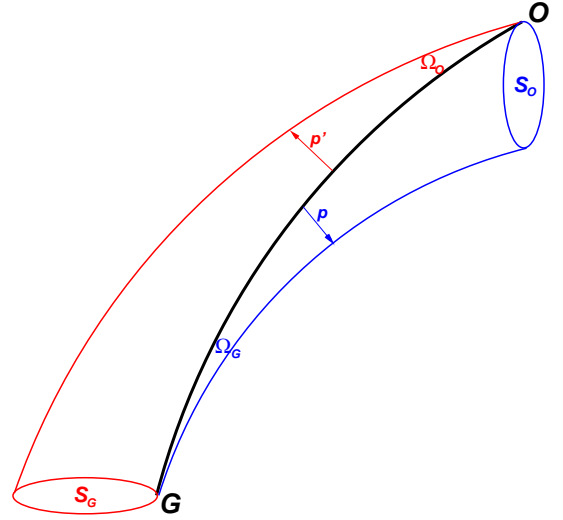}
\end{center}
\caption{Deviation of light rays around the null curve GO linking an observed galaxy with the observer,
  after \cite[][Fig.~7]{Ellis2009}, with several differences in notation. Quantities at
the emission event (i.e. at G) are labelled with G, while quantities at the observation event (i.e. at O) are labelled with $O$.
The connecting vector between the light curves that diverge towards the observer is $p^\alpha$ and the cross section of the diverging bundle at the observation event is $S_{\mathrm O}$ (the observed flux and hence luminosity distance depends on $S_{\mathrm O}$). The connecting vector between the light curves that converge towards the observer is $p'^\alpha$ and the rate of convergence is proportional to $\Omega_{\mathrm O}$ (the observed angular size and hence angular diameter distance depends on $\Omega_{\mathrm O}$).}
\label{fig-reciprocity}
\end{figure}

 Let $k^\alpha$ be a tangent vector of the light curve and let $p^\alpha$ be  a connecting vector for $k^\alpha$.
By definition, a connecting vector is dragged along a congruence of curves associated with and generated by the vector field tangent to the curve, so we have
\begin{equation}
\mathcal{L}_{\mathrm{\mathbf{k}}}\mathrm{\mathbf{p}}=\left[\mathrm{\mathbf{k}},\mathrm{\mathbf{p}}\right]=-\mathcal{L}_{\mathrm{\mathbf{k}}}\mathrm{\mathbf{p}}=p^\alpha_{,\beta} \,k^\beta-k^\alpha_{,\beta} \,p^\beta=0,
\end{equation}
from which follows
\begin{equation}
    k^\beta \nabla_\beta \, p^\alpha
    =
k^\alpha_{;\beta} \,p^\beta+T^\alpha_{\phantom{\alpha}\gamma\beta} \,p^\gamma k^\beta.
\label{derconn}
\end{equation}
Following the steps in \cite{carroll_2019}
we find that $p^\alpha$ obeys the following
equation
\begin{equation}\begin{split}
\frac{\diffD^2 p^\gamma}{\diffD s^2}&= k^\alpha \nabla_\alpha\!\left(k^\beta \nabla_\beta \,p^\gamma\right)=
G^\gamma_{\phantom{\alpha}\sigma\alpha\beta}\,k^\sigma k^\alpha p^\beta + \\
&\;+
k^\alpha k^\beta \nabla_\alpha\!\left(T^\gamma_{\phantom{\alpha}\alpha\beta}\,p^\alpha \right)
+ p^\beta \nabla_\beta \left( T_{\mu\nu}^{\phantom{\mu\nu}\gamma} k^\mu k^\nu \right).
\label{e-p-alpha-deviation}
\end{split}\end{equation}
For a totally antisymmetric torsion
(see \Sect\ref{s-null-curves})
the final term on the right-hand side of the final line here
in \eqref{e-p-alpha-deviation}
is zero.
The tensor $G^\gamma_{\phantom{\alpha}\sigma\alpha\beta}$
\begin{equation}\begin{split}
G^\gamma_{\phantom{\alpha}\sigma\alpha\beta}&=R^\gamma_{\phantom{\alpha}\sigma\alpha\beta}+\frac{1}{2}\left(\nabla_\alpha T^\gamma_{\phantom{\alpha}\sigma\beta}-\nabla_\beta T^\gamma_{\phantom{\alpha}\sigma\alpha}\right)+\\
&\,-\frac{1}{4}\left(T^\gamma_{\phantom{\alpha}\rho\alpha}T^\rho_{\phantom{\alpha}\sigma\beta}-T^\gamma_{\phantom{\alpha}\rho\beta}T^\rho_{\sigma\alpha}\right)-\frac{1}{2}T^\gamma_{\phantom{\alpha}\sigma\rho}T^\rho_{\phantom{\alpha}\alpha\beta}
\label{riemann2}\end{split}\end{equation}
is the generalised Riemann tensor, which is antisymmetric in the first two and last two indices, and $R^\gamma_{\phantom{\alpha}\sigma\alpha\beta}$ is the Riemann tensor defined in terms of the symmetric part of the connection.

The reciprocity theorem links two lengths, usually called
\enquote*{distances}: the observer's
\enquote*{angular diameter distance} $r_{\mathrm O}$,
which is a distance corresponding to light paths tracing backwards in time from O to G,
and the galaxy's \enquote*{angular diameter distance}
$r_{\mathrm G}$ which is a distance corresponding to light paths
tracing forwards in time from G to O. By definition,
  $r_{\mathrm O}$ is
the square root of the cross-sectional area $S_{\mathrm G}$ of a
bundle of light rays from $S_{\mathrm G}$ that converge at O
in a one-steradian solid angle (cf.
Fig.~\ref{fig-reciprocity}); whereas the galaxy's angular diameter
distance $r_{\mathrm G}$  is the square root of
the cross-sectional area $S_{\mathrm O}$ of a one-steradian bundle of
diverging light rays from the galaxy G that arrive at $S_{\mathrm O}$  (cf.
Fig.~\ref{fig-reciprocity}). That is,
\begin{align}
  r_{\mathrm G}^2 := {S_{\mathrm O}}/{\Omega_{\mathrm G}} \,, 
  \quad\quad
  r_{\mathrm O}^2 := {S_{\mathrm G}}/{\Omega_{\mathrm O}} \,.
  \label{e-defn-ang-diam-distances}
\end{align}
We consider a null curve GO that links the observed galaxy with the
observer. We then consider how geodesics deviate from GO. Considering
the connecting vector $p$ will give rise to the galaxy
angular diameter distance
$r_{\mathrm G}$
and
considering the connecting vector $p'$ will give rise to the
observer's angular diameter distance
$r_{\mathrm O}$.

Taking the difference between the geodesic deviation equations leads to:
\begin{equation}\begin{split}
    &p'^\gamma \frac{\diffD^2 p_\gamma}{\diffD s^2} - p^\gamma \frac{\diffD^2 p'_\gamma}{\diffD s^2}=\\ 
    &=\vphantom{\frac{1}{2}}R_{\gamma\sigma\alpha\beta}\,k^\sigma k^\alpha p^\beta p'^\gamma -R_{\gamma\sigma\alpha\beta}\,k^\sigma k^\alpha p'^\beta p^\gamma+\\
    &\,  +\frac{1}{2}\nabla_\alpha T_{\gamma\sigma\beta}\,k^\sigma k^\alpha p^\beta p'^\gamma      -\frac{1}{2}\nabla_\alpha T_{\gamma\sigma\beta}\,k^\sigma k^\alpha p'^\beta p^\gamma+\\
    &\,-\frac{1}{2}\nabla_\beta T_{\gamma\sigma\alpha}\,k^\sigma k^\alpha p^\beta p'^\gamma+\frac{1}{2}\nabla_\beta T_{\gamma\sigma\alpha}\,k^\sigma k^\alpha p'^\beta p^\gamma+ \\
    &\,-\frac{1}{4}T_{\gamma\rho\alpha}\,T^\rho_{\phantom{\alpha}\sigma\beta}\,k^\sigma k^\alpha p^\beta p'^\gamma   +\frac{1}{4}T_{\gamma\rho\alpha}\,T^\rho_{\phantom{\alpha}\sigma\beta}\,k^\sigma k^\alpha p'^\beta p^\gamma+ \\
    &\, +\frac{1}{4}T_{\gamma\rho\beta}\,T^\rho_{\phantom{\alpha}\sigma\alpha}\,k^\sigma k^\alpha p^\beta p'^\gamma  -\frac{1}{4}T_{\gamma\rho\beta}\,T^\rho_{\phantom{\alpha}\sigma\alpha}\,k^\sigma k^\alpha p'^\beta p^\gamma  +\\
    &\,-\frac{1}{2}T^\rho_{\phantom{\alpha}\alpha\beta}\,T_{\gamma\sigma\rho}\,k^\sigma k^\alpha p^\beta p'^\gamma       +\frac{1}{2}T^\rho_{\phantom{\alpha}\alpha\beta}\,T_{\gamma\sigma\rho}\,k^\sigma k^\alpha p'^\beta p^\gamma+ \\
    &\,\vphantom{\frac{1}{2}}-k^\sigma k^\alpha \nabla_\alpha\!\left(T_{\gamma\sigma\beta}\, p^\beta\right)p'^\gamma+      k^\sigma k^\alpha \nabla_\alpha\!\left(T_{\gamma\sigma\beta}\, p'^\beta\right)p^\gamma+ \\
    &\,\vphantom{\frac{1}{2}}+ p^\beta \nabla_\beta\!\left( T_{\sigma \alpha\gamma}\, k^\sigma k^\alpha \right) p'^\gamma
    -  p'^\beta \nabla_\beta\!\left( T_{\sigma \alpha \gamma}\, k^\sigma k^\alpha \right) p^\gamma.
    \label{ciao}\end{split}\end{equation}

\subsection{Riemannian geometry}

We first start with the case of Riemannian geometry and rederive the standard reciprocity theorem.
This is done in order to show how the theorem is affected
by the presence of torsion. The derivation below follows the steps covered in \cite{Ellis2009}.

When torsion vanishes, and utilising the symmetries of the Riemann tensor, we see that the right hand side vanishes, and we obtain

\begin{equation}
\frac{{\diffD} }{{\diffD} s} \left(
p'^\gamma \frac{{\diffD} p_\gamma}{{\diffD} s} - p^\gamma \frac{{\diffD} p'_\gamma}{{\diffD} s}    \right) = 0,
   \end{equation}
and so because of the symmetry of the connection,
\begin{equation}
 p'^\gamma \frac{{\diffd} p_\gamma}{{\diffd} s} - p^\gamma \frac{{\diffd} p'_\gamma}{\diffd s}
\,  = {\rm constant~along~GO}.
\label{tot}\end{equation}

Since  at O, $p'^\gamma = 0$, and at G, $p^\gamma = 0$, it follows that
\begin{equation}
- \left. p^\gamma \right|_{\mathrm O}  \left. \frac{\diffd p'_\gamma}{\diffd s}  \right|_{\mathrm O} =
 \left. p'^\gamma \right|_{\mathrm G}  \left. \frac{\diffd p_\gamma}{\diffd s}  \right|_{\mathrm G} .
  \label{reci}
  \end{equation}

The next step is to choose the connecting vectors such that
they can be linked to the cross-sectional areas and solid angles.
It is always possible to pick two orthogonal directions of propagation,
which remain orthogonal at the observation event:
\[  \left. \frac{\diffd {p_{1}}_a}{\diffd s} \right|_{\mathrm G} \left.\frac{\diffd p_{2}^a}{\diffd s}  \right|_{\mathrm G} = 0, {\rm ~and~} \left. {p_{1}}_a \right|_{\mathrm O}  \left. p_{2}^a\right|_{\mathrm O}  = 0.
\]
Since they are orthogonal, the product of their magnitudes can be linked to the solid angle and area
\[ \diffd \Omega_{\mathrm G} = \left. \frac{\diffd p_{1}}{\diffd l} \right|_{\mathrm G} \left.\frac{\diffd p_{2}}{\diffd l}  \right|_{\mathrm G}
   {\rm ~and~} \diffd S_{\mathrm G} = \left. p_{1}\right|_{\mathrm O}  \left. p_{2}\right|_{\mathrm O},
\]
where $p_1 = \sqrt{p_1^a p_{1a}}$, $p_2 = \sqrt{p_2^a p_{2a}}$,
$\diffd l = \left. (k_\alpha u^\alpha)\right|_{\mathrm G} \diffd s$, and
$\diffd l = - \left. (k_\alpha u^\alpha)\right|_{\mathrm O} \diffd s$ (the minus sign reflects the fact that the direction towards G is opposite to the direction of light propagation).
Similarly, it is also possible to select
 the connecting vectors $p'$ and the initial directions so that they are orthogonal
\[  \left. \frac{\diffd {p'_{1}}_a}{\diffd s} \right|_{\mathrm O} \left.\frac{\diffd {p'_{2}}^a}{\diffd s}  \right|_{\mathrm O} = 0 ~~{\rm and}~~ \left. {p'_{1}}_a \right|_{\mathrm G}  \left. {p'_{2}}^a\right|_{\mathrm G}  = 0, \]
 and so
\[
\diffd \Omega_{\mathrm O} = \left. \frac{\diffd p'_{1}}{\diffd l} \right|_{\mathrm O}  \left. \frac{\diffd p'_{2}}{\diffd l}\right|_{\mathrm O}
{\rm ~and~}
 \diffd S_{\mathrm O} =  \left. p'_{1} \right|_{\mathrm G} \left.   p'_{2} \right|_{\mathrm G}
\]
The condition (\ref{reci}) implies that it is also possible to
select $p$  and $p'$ such that they are co-linear and,
consequently, from  (\ref{reci}) it follows that \cite{PK2006}
\begin{equation}\begin{split}
  & \diffd S_{\mathrm G} \,\diffd\Omega_{\mathrm O} =
  \left( \left. p_{1}\right|_{\mathrm O}  \left. p_{2}\right|_{\mathrm O}, \right)
  \left( \left. \frac{\diffd p'_{1}}{\diffd l} \right|_{\mathrm O}  \left. \frac{\diffd p'_{2}}{\diffd l}\right|_{\mathrm O}  \right)  = \\
  &\, =   \diffd S_{\mathrm O} \,\diffd\Omega_{\mathrm G}  \frac{\nu_{\mathrm G}^2}{\nu_{\mathrm O}^2}
\end{split}\end{equation}
and thus
\begin{equation}
  r_{\mathrm G} = r_{\mathrm O} (1+z) \,.
  \label{e-rG-rO-Riem}
\end{equation}

\subsection{Totally antisymmetric torsion}

If torsion is totally antisymmetric,
then using the (anti)symmetries of the torsion tensor we obtain
\begin{equation}
    p'^\gamma \frac{{\diffD}^2 p_\gamma}{{\diffD} s^2} - p^\gamma \frac{{\diffD}^2 p'_\gamma}{{\diffD} s^2}= -  k^\alpha \nabla_\alpha \left(
     T_{\gamma\sigma\beta}\, k^\sigma p^\beta p'^\gamma \right),
   \end{equation}
which can be rewritten as
\begin{equation}
\frac{{\diffD} }{{\diffD} s} \left(
p'^\gamma \frac{{\diffD} p_\gamma}{{\diffD} s} - p^\gamma \frac{{\diffD} p'_\gamma}{{\diffD} s}    +T_{\gamma\sigma\beta}\, p^\beta k^\sigma p'^\gamma \right) = 0 .
\label{reci2}
   \end{equation}
By means of equations \eqref{e-p-alpha-deviation} and \eqref{derconn}, this can be reworked as follows
\begin{equation}\begin{split}
&p'^\gamma \frac{{\diffD} p_\gamma}{{\diffD} s} - p^\gamma \frac{{\diffD} p'_\gamma}{{\diffD} s}    +T_{\gamma\sigma\beta}\, p^\beta k^\sigma p'^\gamma \\
=&\,p'^\gamma k^\alpha p_{\gamma,\alpha}- p^\gamma k^\alpha p'_{\gamma,\alpha}-T_{\sigma\gamma\beta}\, p^\beta k^\sigma p'^\gamma-T_{\beta\gamma\sigma}\, p^\beta k^\sigma p'^\gamma.
\end{split}\end{equation}
Because of the (anti)symmetries of the torsion tensor, the last two terms cancel out. Thus the same result as in the standard case (eq. \eqref{tot}) is recovered
\begin{equation}\begin{split}
& p'^\gamma \frac{\diffd p_\gamma}{\diffd s} - p^\gamma \frac{\diffd p'_\gamma}{\diffd s} = {\rm constant~along~GO}.
\end{split}\end{equation}

As above, at O, $p'^\gamma = 0$, while at G, $p^\gamma = 0$,
so we recover the same results as in the case of Riemannian geometry, that is, we obtain
\eqref{reci}. Consequently, in the case of
totally antisymmetric torsion,
the reciprocity theorem still holds \cite{2017PhRvD..95f1501S},
 and thus
 \[ r_{\mathrm G} = r_{\mathrm O} (1+z). \]

\subsection{General case with torsion}

In the general case, the  difference  between  the  geodesic  deviation equations (\ref{ciao}), can be written as in the case of the totally antisymmertic torsion  (\ref{reci2}), with the remaining terms which we combine into a function  $j(T)$

\begin{equation}
 \frac{{\diffD} }{{\diffD} s} \left(
p'^\gamma \frac{{\diffD} p_\gamma}{{\diffD} s} - p^\gamma \frac{{\diffD} p'_\gamma}{{\diffD} s}    +T_{\gamma\sigma\beta} p^\beta k^\sigma p'^\gamma \right) = j(T) \,.
   \label{gedevt}
   \end{equation}
For the homogeneous and isotropic background model of Ref.~\cite{Kranas2019},
$j(T) \sim \phi^2$ and thus $j>0$, hence
\begin{equation}
 \left. p^\gamma \right|_{\mathrm O}  \left. \frac{\diffd p'_\gamma}{\diffd l}  \right|_{\mathrm O} =
 \left. p'^\gamma \right|_{\mathrm G}  \left. \frac{\diffd p_\gamma}{\diffd l}  \right|_{\mathrm G} (1+z)
 + \int\limits_{\mathrm G}^{\mathrm O} {\rm d}s \, j(T),
     \end{equation}
Consequently,
\begin{equation}
r_{\mathrm G}^2 = r_{\mathrm O}^2 (1+z)^2 (1 + g(T)),
\label{recitorsion}
\end{equation}
where $g(T)$ is of the order of $\phi^2$ and thus positive,
which implies that in the presence of torsion,
$r_{\mathrm G}$ is greater than $r_{\mathrm O}(1+z)^2$.

\section{Cosmic duality relation}

Given the luminosity $L$ of the source galaxy at G and the observed radiation flux $f$,
the luminosity distance $D_{\mathrm L}$ is defined by
\begin{equation}
  D_{\mathrm L} := \left(\frac{L}{4\pi f}\right)^{1/2},
  \label{e-DL-defn}
\end{equation}
so that, by definition, flux $f$ drops as the square of the luminosity \enquote*{distance} $D_{\mathrm L}$.
The angular diameter distance is defined as above
\begin{equation}
  D_{\mathrm A}^2 = \frac{S}{\Omega} = r_{\mathrm O}^2 \,.
  \label{e-DA-defn}
\end{equation}

The luminosity distance is related to $r_{\mathrm G}$ as defined above.
By considering flux, we remind the reader in \Sect\ref{s-GR-duality-relation} of the GR derivation
of the GR cosmic duality relation that
relates $D_{\mathrm A}$ to $D_{\mathrm L}$:
\begin{equation}
  \frac{ D_{\mathrm L}}{D_{\mathrm A}} = (1+z)^2 \,.
  \label{e-GR-duality-relation-expected}
\end{equation}

\subsection{Flux and its evolution in GR} \label{s-GR-duality-relation}

The energy--momentum tensor of radiation can be approximated as
above in \eqref{em-rad},
\begin{equation}
  T^{\alpha\beta}_{\mathrm{rad}} =  A^2 k^\alpha k^\beta \,.
  \label{e-radiation-Emom}
\end{equation}
The flux of radiation as measured by an observer with four-velocity $u^\alpha$ is
\begin{equation}
f = T_{\alpha \beta}^{\mathrm{rad}} \, u^\alpha u^\beta = \nu^2 A^2.
\end{equation}
At the source, the flux can be linked to the luminosity of the source
\begin{equation}
 f_{\mathrm G} =
  \frac{L}{4 \pi r^2_{\mathrm G}} =
 \frac{L}{4 \pi} \frac{\Omega_{\mathrm G}}{ S_{\mathrm G}}
  \label{e-amplitude-expression}
\end{equation}
where $\Omega_{\mathrm G}$ is the (small) solid angle and $S_{\mathrm G}$ is the cross section of the light bundle (cf. Fig. \ref{fig-reciprocity} and eq. (\ref{e-defn-ang-diam-distances})).

In GR, the product of $A^2$ and $S$ is constant along the light curve, i.e.
$A_{\mathrm G}^2 S_{\mathrm G} = A_{\mathrm O}^2 S_{\mathrm O}$ \cite{1966ApJ...143..379K,Ellis2009}.
Consequently, the product of $A^2$ and $S$  at the observer can be replaced by the product at the emitting galaxy. Subsequently, the flux--luminosity relation
\eqref{e-amplitude-expression}
and the definition of $r_{\mathrm G}$ \eqref{e-defn-ang-diam-distances}
yield
\begin{equation}
f_{\mathrm O} = \nu_{\mathrm O}^2 A_{\mathrm O}^2 =
\nu_{\mathrm O}^2
\frac{A_{\mathrm G}^2 S_{\mathrm G}}{S_{\mathrm O}} =
\frac{\nu_{\mathrm O}^2}{\nu_{\mathrm G}^2} \frac{L}{4\pi r_{\mathrm G}^2} \,.
\label{e-flux-at-Observer}
\end{equation}

Applying the definition of $D_{\mathrm L}$
\eqref{e-DL-defn},
adopting the definition of redshift (i.e.
$\nu_{\mathrm O}/ \nu_{\mathrm G} = 1+z)$,
using the Riemannian $r_{\mathrm G}$--$r_{\mathrm O}$ relation \eqref{e-rG-rO-Riem},
and dropping the prefix O,
we obtain the standard GR result
\begin{equation}
  D_{\mathrm L}^2 = \frac{L}{4\pi f} = (1+z)^2 r_{\mathrm G}^2 = (1+z)^4 D_{\mathrm A}^2 \,.
  \label{e-GR-FLRW-duality-relation}
\end{equation}

\subsection{Flux and its evolution in the presence of torsion}\label{flux}

In the presence of torsion the above relation is modified. Assuming
that photons can be treated as a small perturbation and decoupled from
matter so that they do not affect the overall geometry of the
spacetime and assuming that the Einstein equations
$G^{\mu \nu} = 8 \pi T^{\mu \nu}$ still hold (as for example in the case of the Einstein-Cartan gravity), we have
\begin{equation}
 \nabla_\nu  T^{\mu \nu}_{{\rm mat}} = \frac{1}{8 \pi}  \nabla_\nu G^{\mu \nu}
  {\rm~~and~~}
\nabla_\nu  T^{\mu \nu}_{{\rm rad}} = 0.
\label{em-temsor}
\end{equation}
Contracting the second of these two
equations with the observer's four-velocity $u_\mu$
and using
\eqref{e-radiation-Emom},
we obtain
\begin{equation}
(A^2)_{;\alpha} k^\alpha  = - A^2 k^\alpha{}_{;\alpha}   + A^2 \nu T,
\end{equation}
where $T$ is as defined just below eq.~\eqref{redshiftpropagation},
i.e. $T = T_{ \alpha  \beta \sigma}\,  u^\alpha n^\beta u^\sigma
+ T_{ \alpha  \beta \sigma}\,   n^\alpha n^\beta u^\sigma $.

Thus, writing down the evolution of $A^2 S$, we obtain
\begin{equation}
\frac{1}{ A^2 S } \frac{{\diffD} }{{\diffD} s} (A^2 S) =\nu T.
\end{equation}
Writing the integral of the RHS as a function $b$ defined as
\begin{equation}
   b :=  \int\limits_{\mathrm G}^O {\rm d}t \, T,
\label{beint}
\end{equation}
and by assuming that $b$ is small (${\rm e}^b \approx 1 +b $), the observed flux is

\begin{equation}
  f_{\mathrm O} = \nu_{\mathrm O}^2 A_{\mathrm O}^2 =
  \nu_{\mathrm O}^2
  \frac{A_{\mathrm G}^2 S_{\mathrm G}}{S_{\mathrm O}} (1+b) =
  \frac{\nu_{\mathrm O}^2}{\nu_{\mathrm G}^2} \frac{L}{4\pi r_{\mathrm G}^2} (1+b)  \,.
  \label{e-flux-at-Observer-torsion}
\end{equation}
Using (\ref{recitorsion}), we obtain
the cosmic duality relation affected by the presence of torsion is
\begin{equation}
    D_{\mathrm L}^2 = (1+z)^4  \frac{(1+g)}{(1+b)} D_{\mathrm A}^2.
\label{cosmic-duality-torsion}
\end{equation}
We adopt the commonly used
notation for empirical estimates of the deviation of the distance duality relation
\begin{equation}
  \eta :=  \frac{D_{\mathrm L}}{(1+z)^2 D_{\mathrm A}} - 1,
  \label{e-eta-defn}
\end{equation}
which, if torsion is weak (i.e. both $|b|$ and $|g|$ are much
smaller than unity), yields
\begin{equation}
 \eta = \sqrt{(1+g)(1-b)} -1  \approx \frac{1}{2} (g-b).
 \label{etatorsion}
\end{equation}
The parameter $g$ describes the effect of torsion on the cross section of the bundle, and the parameter $b$ describes the effect of torsion on the flux.

\section{Distinguishable cosmological signatures}\label{cosimp}

\subsection{Homogeneous and isotropic models}

\begin{table}
  \begin{center}
    \caption{Parametrisation of torsion and deviation from the distance duality relation}
    \label{tab1}
    \begin{tabular}{l|l}
  torsion  $\phi \quad$   & \quad distance duality $\eta$  \\ \hline
  $\phi = \eta_0 H_0 \quad$ & $\quad \eta =  \eta_0 \, z$ \rule{0ex}{2.5ex}
  \\
  $\phi = \eta_0 Ha \quad$ & $\quad \eta = \eta_0 \dfrac{z}{1+z}$   \\
  $\phi = \eta_0 H \quad$ & $\quad \eta = \eta_0 \ln (1+z)$  \rule{0ex}{2.5ex}
    \end{tabular}
  \end{center}
\end{table}

In this section we assume homogeneity and isotropy and use the models of \cite{Kranas2019}.
Let us assume that the metric of a universe model is
described by the FLRW line element
\begin{equation}
    \diffd s^2 = -{\diffd}t^2 + a^2 \left(
    \frac{{\diffd}r^2}{1-kr^2} + r^2 {\diffd} \vartheta^2 + r^2 \sin \vartheta^2 {\diffd} \varphi^2 \right),
    \label{flrwmetric}
\end{equation}
where $a$ is dimensionless, $r$ has units of length,
and at the equator $1=kr^2$ in the spherical case, the algebraic singularity in the expression is
replaced by $g_{rr} = \lim_{r' \rightarrow r} \left[{a^2}/({1-kr'^2})\right]$.
Considering null curves $\diffd s=0$, the redshift formula is
\begin{equation}
  1+z = \frac{1}{a(t)}.
  \label{Apar}
\end{equation}
Thus, for homogeneous and isotropic models the integrand on the RHS of
(\ref{ltb}) needs to be ${\dot{a}}/{a}$. Indeed,  for homogeneous and isotropic models with the FLRW metric
and non-zero torsion \cite{Kranas2019}
\begin{equation}
T_{ \alpha \beta \sigma}\,  u^\alpha n^\beta u^\sigma= 0 ~~{\mathrm{and}}~~ T_{ \alpha \beta \sigma}\,  n^\alpha n^\beta u^\sigma = -2\phi,
\label{homtor}
\end{equation}
and the integrand is
\begin{equation}
 \frac{1}{3} \Theta - 2 \phi \equiv H = \frac{\dot{a}}{a},
 \label{Hpar}
\end{equation}
which is consistent with the definition of $H$ obtained by comparison of the Einstein--Cartan equations with the Raychaudhuri equation \cite{Kranas2019}.

The angular diameter distance is by definition $D_{\mathrm A}^2 = S/\Omega$, cf. \eqref{e-DA-defn}.
Based on the metric (\ref{flrwmetric}), it is
\begin{equation}
D_{\mathrm A} = a(t) D,
\label{flrwda}
\end{equation}
where $D$ is the transverse comoving distance (comoving one-radian arc length) given by
\begin{align}
  D &=
  \lim_{k' \rightarrow k}
  \frac{1}{\sqrt{k'}} \,
  \sin\left( \sqrt{k'} \int_{\mathrm O}^G  \frac{{\rm d} r}{\sqrt{1-k'r^2}} \right) \nonumber \\
  &= \lim_{k' \rightarrow k}
  \frac{1}{\sqrt{k'}} \,
  \sin\left( \sqrt{k'}
  \int_{\mathrm G}^O  \frac{ {\rm d} t}{a} \right) \,,
  \label{Dpar}
\end{align}
where the hyperbolic case ($k<0$) yields a $\sinh$ dependence ($\sin  i x = i \sinh x$).
Although the evolution of the scale factor $a(t)$ is affected by torsion, and consequently the transverse comoving distance, the structure  of the relation between the transverse comoving distance and the angular diameter distance is not affected by the presence of torsion, i.e. torsion does not explicitly appear in this relation.
This is in contrast to the luminosity distance, where torsion explicitly affects the flux formula, and thus the cosmic duality relation (\ref{cosmic-duality-torsion}),
where by the definition of $\eta$ \eqref{e-eta-defn},
\begin{equation}
    D_{\mathrm L} = (1+z)^2 \, D_{\mathrm A} \, (1+\eta).
    \label{flrwdl}
\end{equation}
Consequently, we may say that torsion can affect the luminosity distance
in two ways. The first factor is via the
scale factor evolution $a(t)$ that solves the
Einstein--Cartan equations, which generalise beyond the GR case.
Non-GR effects on the evolution of $a(t)$
would affect the angular diameter distance
dependence on $a(t)$ in \eqref{Dpar} and \eqref{flrwda}.
The second factor affecting the luminosity
distance is by torsion affecting the flux itself via
(\ref{e-flux-at-Observer-torsion}), which follows through to
\eqref{flrwdl}, showing the dependence of the luminosity distance on the
distance duality deviation parameter $\eta$.

The evolution of this type of
universe is governed by the Einstein--Cartan equations \cite{Kranas2019}
\begin{eqnarray}
 && \dot{\Theta} = -\frac{1}{3} \Theta^2 - \frac{1}{2} \kappa \rho + \Lambda + 2 \Theta \phi \label{dyneqs1} \\
 && \dot{\rho} = - \Theta  \rho + 4 \phi \left( \rho + \kappa^{-1} \Lambda  \right),
 \label{dyneqs2}
\end{eqnarray}
where $\kappa := 8 \pi G$, $\rho$ is the density of matter, $\phi$ is torsion,
$\Lambda$ is a cosmological constant,
and $\Theta$ is the expansion rate, related to the scale factor via (\ref{Hpar}).
\begin{equation}
    \dot{a} = \frac{1}{3} \Theta a - 2 \phi a.
    \label{dyneqs3}
\end{equation}
Just as in the case of FLRW models, the parameters $\Omega$ can also be introduced for a homogeneous and isotropic model with torsion \cite{Kranas2019}
\begin{eqnarray}
&& \Omega_m = \frac{\kappa \rho}{3 H^2}, \nonumber \\
&& \Omega_k = - \frac{k}{a^2 H^2},  \nonumber  \\
&& \Omega_\Lambda = \frac{\Lambda}{3H^2},  \nonumber  \\
&& \Omega_\phi = -4 \left( 1 + \frac{\phi}{H} \right) \frac{\phi}{H},  \label{comegas}
\end{eqnarray}
where the last equations above follows from (\ref{dyneqs1}), (\ref{dyneqs3}), and
\begin{equation}
    \Omega_m + \Omega_k + \Omega_\Lambda + \Omega_\phi  = 1.
\end{equation}

\subsection{Distance duality within the approximation of weak torsion}\label{torparam}

For a given model of torsion $\phi(t)$,  one can integrate (\ref{beint}) as well as (\ref{gedevt}) and get the exact formula for the deviation from the distance duality (\ref{etatorsion}).
The approximation of weak torsion means that the contribution
of torsion to the reciprocity relation (\ref{recitorsion}) is assumed to be small compared to the contribution
of torsion to the flux relation (\ref{e-flux-at-Observer-torsion}).
For homogeneous and isotropic models with torsion, $b \sim \phi$ and $g \sim \phi^2$. Thus, in the limit of weak torsion, using (\ref{homtor}),  the deviation from the distance duality (\ref{etatorsion}), can be further approximated as
\begin{equation}
    \eta \approx -\frac{1}{2} b = \int\limits_{\mathrm G}^O {\rm d}t \, \phi.
    \label{etaapproximation}
\end{equation}
Since torsion has units of the expansion rate, most common parametrisations of torsion are based on the relation \cite{Kranas2019,2019EPJC...79..950P}
\[ \phi \sim H. \]
Assuming some standard parametrisations, we can integrate \eqref{etaapproximation}
to find the formula for the deviation from the distance  duality in the limit of weak torsion

\begin{itemize}
\item $\phi = \eta_0  H_0$

  In this case, torsion is assumed to be constant, which would not
  seem to be a reasonable assumption unless only local
  observations are considered. Assuming that we indeed deal only with
  local observations $\Delta t \approx z/H_0$, we obtain
  \begin{equation}
    \eta \approx \eta_0 z\,.
    \label{e-eta-const}
  \end{equation}
\item $\phi = \eta_0  H$

  In this case, torsion is linked to the expansion rate of the
  universe model. Integrating (\ref{etaapproximation}) yields
  \begin{equation}
    \eta \approx \eta_0 \ln (1+z)\,.
    \label{e-eta-expansion}
  \end{equation}
\item $\phi = \eta_0  H a^n $

  In this case, torsion is expressed in terms of the expansion
  rate and the scale factor (cf. \cite{2019EPJC...79..950P}).
  Integrating (\ref{etaapproximation}) gives
  \begin{equation}
    \eta \approx \eta_0 \frac{1}{n} \frac{ (1+z)^{n} -1 }{(1+z)^{n}} \,.
  \end{equation}
  The limiting case of $n=0$ is considered above. The case of $n=1$ leads to
  \begin{equation}
    \eta \approx \eta_0 \frac{z}{1+z}\,,
    \label{e-eta-evolving}
  \end{equation}
  which is another commonly used empirical parametrisation for the
  departure from the GR distance duality relation, $\eta \sim  z/(1+z)$.
\end{itemize}

\begin{figure}
\begin{center}
  \includegraphics[scale=0.65]{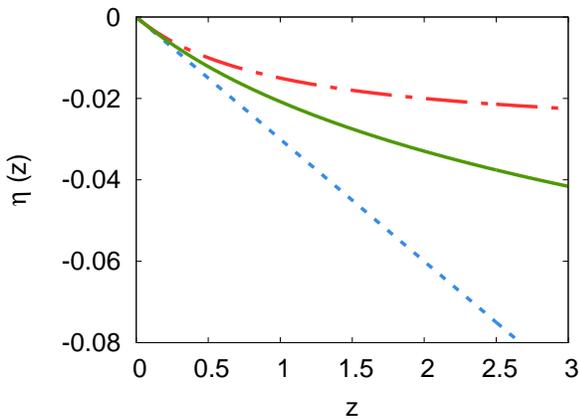}
\end{center}
\caption{The deviation $\eta$ \protect\eqref{e-eta-defn}
  of the distance duality relation evaluated
  in the limit of weak torsion (\ref{etaapproximation}),
  as a function of redshift $z$ for three different torsion models
  with $\eta_0 = -0.03$.
  Bottom curve (dotted blue) \eqref{e-eta-const};
  middle curve (solid green) \eqref{e-eta-expansion};
  top curve (dot-dashed red): \eqref{e-eta-evolving}.}
\label{fig-eta}
\end{figure}

A summary of these parametrisations is presented in Table~\ref{tab1},
and an example is presented in Fig.~\ref{fig-eta} with $\eta_0=-0.03$.

\begin{figure}
\begin{center}
\includegraphics[scale=0.65]{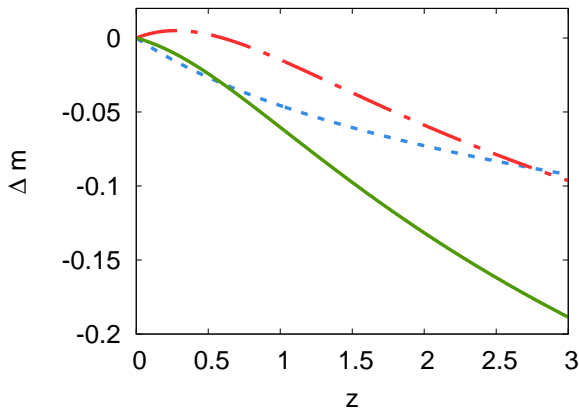}
\end{center}
\caption{Differential effects on the apparent magnitude of a standard candle induced
  by the torsion model given in \protect\eqref{e-tlcdm-torsion-model} in comparison
  to a reference $\Lambda$CDM model.
  Top curve (dot-dashed red): $5 \log_{10} \Delta_D$;
  middle curve (dotted blue): $5 \log_{10} (1+\eta) $;
  bottom curve (solid green): $5 \log_{10} (1+\eta) +
  5 \log_{10} \Delta_D$. The bottom line shows the total change of magnitude within the approximation of weak torsion and with $\eta_0 = -0.03$. As seen a change of magnitude of order of $\Delta m = 0.1$ is small (compared to for example the scatter of supernovae data) but still it is not negligibly small, and could be mistaken (for example) for the evolution of sources, with sources misinterpret to be brighter at larger redshift.
  \label{fig-deltam}}
\end{figure}

\subsection{Impact on standard-candle distance moduli} \label{s-impact-m-M}

As stated above, torsion affects cosmological observations in
two ways: it modifies scale factor evolution $a(t)$,
which affects the angular diameter distance
((\ref{flrwda}) and (\ref{Dpar}));
and it affects the flux via $\eta$ in
(\ref{flrwdl}).

To estimate the significance of these effects, let us compare two
models: (1) a $\Lambda$CDM model with
current-epoch density parameters
$\Omega_{\mathrm m0} = 0.3$, $\Omega_{\mathrm \Lambda0} = 0.7$, and
Hubble constant $H_0 = 70$~{\kmsMpc};
(2) a \enquote*{T$\Lambda$CDM} model which is the same as the
$\Lambda$CDM model, but has an empirical torsion model
\begin{equation}
  \phi := \eta_0 H\,, \quad\quad \eta_0 := -0.03 \,.
  \label{e-tlcdm-torsion-model}
\end{equation}
As follows from (\ref{comegas}) with torsion present, the model is no longer flat as $\Omega_k \approx 0.1$.

The evolution equations (\ref{dyneqs1})--(\ref{dyneqs3}) are
solved starting from the present epoch $t_0$ and traced back in
time.  The initial conditions are:
\[ a(t_0) = 1,~~\rho(t_0) = \Omega_{\mathrm m 0} \frac{3 H_0^2}{8\pi G},~~\Theta(t_0) = 3 H_0 + 6\phi_0.\]

The impact on observations is measured in terms of the change of the
bolometric apparent magnitude
\begin{eqnarray}
 m \, &=& -2.5 \log_{10} f_{\mathrm O} + {\mathrm{const}} \nonumber \\
&=& 5 \log_{10} \left(D_{\mathrm L}/{1\mathrm{~Mpc}}\right) + 25 + M ,
\label{magnitude}
\end{eqnarray}
where the constant relates to the zero-point of the magnitude system and
the choice of units for expressing the flux $f_{\mathrm O}$,
$M$ is the absolute magnitude of an observed extragalactic object.
We assume that the object is a standard candle, i.e.
it emits the same spectral energy distribution
independent of the age of the universe at the time of emission and the lookback time, so the $K$-correction \cite{Wirtz1918} is zero.

Since we have adopted an FLRW metric, a fixed redshift $z$ implies that the scale factor
$a$ is fixed via the usual relation \eqref{Apar}. The scale factor evolution is,
in general, affected by torsion, so the inverse function $t(a)$ (over an
always-expanding range of epochs $a$) will modify the age of the universe at
the chosen redshift. Nevertheless,
since we are considering a standard candle, there is no effect on
the object's spectral energy distribution.
Thus, for fixed $z$,
the difference in magnitude when comparing the two different models is
\begin{equation}
  \Delta m = 5 \log_{10} (1+\eta) +
  5 \log_{10} \Delta_D \,,
\label{deltams}
\end{equation}
 where $\Delta_D$ is the change in
the comoving transverse distance induced by the effect of torsion
on the expansion history,
\[
\Delta_D := \frac{D}{D_{\Lambda\mathrm{CDM}}}\,,
\]
where $D$ is the transverse comoving distance given by (\ref{Dpar}).

The contributions of $\eta$ and $\Delta_D$ to apparent magnitude are compared
in Fig.~\ref{fig-deltam}.
If an effect such as that shown in Fig.~\ref{fig-deltam} is not accounted for,
then cosmological observations interpreted in terms of GR models will yield
anomalous results.
Thus, torsion can
impact the way that supernovae of type Ia
are calibrated and inferences made regarding the value of
the Hubble constant \cite{2017JCAP...01..038C,2019MNRAS.490.2948D}.
For example, as seen from Fig.~\ref{fig-deltam}, not accounting for torsion can lead to a change in apparent magnitude. In the particular example presented in
Fig.~\ref{fig-deltam}, the change is negative, meaning that the sources would appear brighter than expected. This could be misinterpreted
in terms of the evolution of the sources
(i.e. supernovae being brighter in the past than at the present) if torsion
were not taken into account.

\begin{figure}
\begin{center}
\includegraphics[scale=0.65]{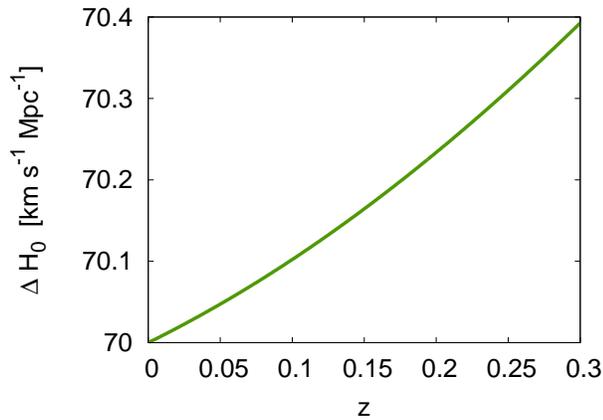}
\end{center}
\caption{
The impact of cosmological torsion on the inferred value of $H_0$ as a function of redshift of the data used, i.e. if one uses data at $z \approx 0.3$, then the inferred value of $H_0$ is greater by $0.4$ km s$^{-1}$ Mpc$^{-1}$ compared to $H_0$
inferred from the data at $z\approx 0$.
The effect of torsion on the luminosity distance was evaluated within the approximation of weak torsion, i.e.
\protect\eqref{etaapproximation} with \protect\eqref{e-tlcdm-torsion-model} and $\eta_0=-0.03$.}
\label{fig-deltah}
\end{figure}

Another example of how not accounting for torsion could lead to misinterpretation of cosmological observations is presented in
Fig.~\ref{fig-deltah}. The change of the magnitude (\ref{deltams})
could be misinterpreted, not in terms of the absolute magnitude $M$,
but rather in terms of the change of the value of the
Hubble constant inferred from observations. Following from  (\ref{magnitude}),
\begin{equation}
    \Delta m = - 5 \log_{10} \left(1 + \frac{\Delta H_0}{H_0}\right) \,.
    \label{dehz}
\end{equation}
Thus, the change of apparent magnitude produced by the presence of torsion could be misinterpreted in terms of an apparent change of the expansion rate. This is presented in Fig.~\ref{fig-deltah}.
The results presented in Fig.~\ref{fig-deltah} show that if we were to use data from around $z \approx 0.2$, then the inferred value $H_0$ would be about $0.2$ km s$^{-1}$ Mpc$^{-1}$ greater than the value of $H_0$ inferred from the data at $z\approx 0$. A phenomenon that is qualitatively similar to this, but stronger and with a sign reversal, is the \enquote*{Hubble bubble} \cite{2007ApJ...659..122J}, which remains present in recent data,
e.g., \cite{2016ApJ...826...56R,2019MNRAS.490.4715S}.
While several other effects could result in a similar apparent change of the Hubble constant \cite{2017JCAP...01..038C,2019MNRAS.490.2948D}, it is interesting that torsion could lead to such changes. Future precise measurements of supernova data at low redshifts could be used to study cosmological torsion.

\begin{figure}
\begin{center}
\includegraphics[scale=0.65]{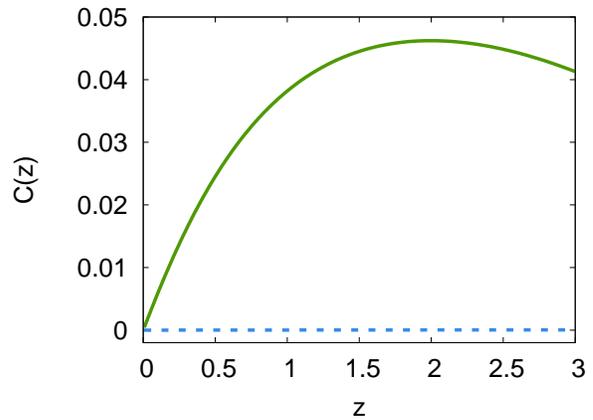}
\end{center}
\caption{The CBL parameter as given by eq. (\protect\ref{CBLfun}).
  If the distance $D$ is obtained from the angular diameter distance
  \protect\eqref{flrwda}, \protect\eqref{Dpar}, then the CBL parameter
  is zero, ${\cal C}(z) = 0$ (flat blue curve).  If the distance is
  inferred from supernovae type Ia distance modulus observations,
  i.e. from the luminosity distance \protect\eqref{flrwdl},
  without accounting for the presence of torsion via
  the parameter $\eta$, then ${\cal C}(z) \ne 0$ (green curve).
  The torsion model here is the same as for the other figures
  \protect\eqref{e-tlcdm-torsion-model}, i.e.
  $\phi = \eta_0 H$, $\eta_0=-0.03$.
  \label{fig-CBL}}
\end{figure}

\subsection{Clarkson--Bassett--Lu function}

The aim of the above considerations was to show that torsion could affect and bias the results of cosmological observations. However, a similar effect could be accommodated by, for example, a dark energy model with an evolving equation of state, or by a model with interaction in the dark sector. How can one tell if indeed the observed signatures are produced by torsion rather than by a dark energy model within general relativity?

A standard dark energy model within GR (no matter how complicated the equation of state is or how complex the interactions between dark matter and dark energy are) will not affect the cosmic duality relation.
Thus, measurement of $D_{\mathrm A}$ and $D_{\mathrm L}$ will show if the parameter
$\eta$ is non-zero.

Another possibility for detecting deviation from Riemannian geometry
is to use the Clarkson--Bassett--Lu (CBL) function ${\cal C}(z)$,
defined in \cite{2008PhRvL.101a1301C}
and derived from the relation between the expansion history and spatial
curvature in FLRW models,
\begin{equation}
{\cal C}(z) = 1 + H^2 (D D'' - D'^2) + H H' D D',
\label{CBLfun}
\end{equation}
where $H$ is given by (\ref{Hpar}) and $D$ is given by (\ref{Dpar}).
In the FLRW case \eqref{flrwmetric}, $C(z) \equiv 0 $ for the Riemannian geometry.
Deviations from zero
are particularly interesting for indicating inhomogeneous GR models
(in which structure formation is taken into account;
\cite[e.g.][]{2009PhRvD..80l3512W,ROB13,2016IJMPD..2530007B,RMBO16Hbg1} and references therein for recent observational
calibrations).
However, deviations from zero
could also indicate that non--pseudo-Riemannian geometry is
required to describe our Universe.
For the models considered here, i.e. models with torsion,
the value of the function $C$ will depend on the type of observations
used to infer the parameters $H$ and $D$.

If the Hubble parameter $H$ is derived from observations that directly use
\eqref{Apar} and \eqref{Hpar}, and if the distance $D$ is
obtained based on the angular diameter distance (\ref{flrwdl}), then
consistency between the expansion history and curvature will
indeed yield ${\cal C}(z) = 0$ for the FLRW metric. However, if the distance
is inferred from supernovae observations, i.e. from luminosity distances
(\ref{flrwdl}), without accounting for the presence of torsion via the
parameter $\eta$, then ${\cal C}(z) \ne 0$ will be inferred.
This is illustrated in Fig.~\ref{fig-CBL}.

Thus, not only should measuring $D_{\mathrm A}$ and $D_{\mathrm L}$ provide constraints
on torsion models, but the
presence of torsion should have an observational signature in the
${\cal C}(z)$ relation if $D$ is inferred from the luminosity
distance (distance modulus).

\subsection{Reproducibility of results}

The code that was used to produce the results reported in this paper is free-licensed software available at \url{https://codeberg.org/bolejko/cosmotor.git}. The code is written in standard fortran (compatible with the f95, f2003 and f2008 standards)
and does not require any additional libraries to run.
The reader is encouraged to run the code to reproduce the results shown here,
and to modify the code and redistribute improved versions to
derive their own results.

The code solves for the transverse comoving distance (\ref{Dpar}) as a function of redshift, together with the evolutionary equations (\ref{dyneqs1})--(\ref{dyneqs3}).
The transverse comoving distance is then used to evaluate relevant observational signatures such as, change of the magnitude as given by (\ref{deltams}).
The code also evaluates the derivatives with respect to redshift of the distance as well as the expansion rate -- these were needed to evaluated the CBL function (\ref{CBLfun}) -- and thus can easily be implemented in other studies of cosmological torsion.

The repository available at \url{https://codeberg.org/bolejko/cosmotor.git} also includes gnuplot scripts to reproduce Figures~3--6. Thus, if the reader compiles and executes the code, they can also use \texttt{gnuplot}\footnote{\url{http://www.gnuplot.info/}} to reproduce exact copies of Figures~3--6.

\section{Conclusions}

The aim of this paper was to consider observational signatures of
spacetime torsion on cosmological scales.  Several cosmological models
with torsion have recently been considered in the literature
\cite{2019EPJC...79..764B,2019EPJC...79..950P,Kranas2019,2020PDU....2700416M}.
The motivation in this paper differs from the approach of these studies.
Instead of model comparison, we focus on observational features
that are most likely to be clearly distinguishable from those of dark energy models, and
which are not reproducible by any dark energy model, i.e. \enquote*{litmus
test} type signatures.

We show  that if torsion is present, then the cosmic duality relation,
parametrised by $\eta$ \eqref{e-eta-defn}, is broken:
\[ \frac{D_{\mathrm L}}{D_{\mathrm A}(1+z)^2} -1= \eta  \ne 0. \]
A special case is for the totally antisymmetric torsion tensor, in which case the
cosmic duality relation holds: $\eta \equiv 0$ \cite{2017PhRvD..95f1501S}.
However, homogeneous and isotropic cosmological models with an
FLRW-type metric do not, in general, have totally antisymmetric torsion
\cite{Kranas2019}, and thus, cosmological models with torsion should,
in general, violate the distance duality relation: $\eta \ne 0$.

Current observational constraints on $\eta$ are not conclusive, and
still consistent with $\eta = 0$ \cite{2017IJMPD..2650097F}.  With
$D_{\mathrm L}$ obtained from supernova observations and $D_{\mathrm A}$
from lensing, one can constrain the parameter $\eta$. However,
$D_{\mathrm A}$ estimates of this sort are strongly dependent
on the density profiles adopted for the lensing analyses.
For $\eta = \eta_0 \ln (1+z)$, the constraints found on $\eta_0$
were $\eta_0 = 0.21^{+0.16}_{-0.19}$ (for a singular isothermal sphere profile),
and $\eta_0 = -0.22^{+0.14}_{-0.20}$ (for a power law profile)
\cite{2017IJMPD..2650097F}.

We found that distance duality violation is not the only
signature of torsion. While less conclusive, as similar effects could
result from other models, low-redshift observations could also be
used to point towards torsion, with small biases in Hubble constant
estimates (Figs~\ref{fig-deltam}, \ref{fig-deltah}) or a non-zero value of
the CBL function ${\cal C}(z)$ if the distances are derived from distance
moduli (Fig.~\ref{fig-CBL}).

\section*{Acknowledgements}

We thank Mariana Jaber for several useful suggestions.
KB acknowledges support from the Australian Research Council through his Future Fellowship FT140101270.
Part of MC's contribution to this work was supported by Universitas Copernicana Thoruniensis in Futuro under NCBR grant POWR.03.05.00-00-Z302/17.
Part of this work was supported by the ``A next-generation worldwide quantum sensor network with optical atomic
clocks'' project, which is carried out within the TEAM IV programme of the
Foundation for Polish Science co-financed by the European Union under the
European Regional Development Fund.
Part of this work was performed under MNiSW grant DIR/WK/2018/12.

\bibliography{cosmotor}
\bibliographystyle{apsrev}

\end{document}